\documentclass{ws-p8-50x6-00}
\begin{document}
\title{Chiral Effective Field Theories with Explicit Spin 3/2 Degrees of
Freedom---A Status Report}
\author{Thomas R. Hemmert}
\address{Physik Department T39, TU M{\" u}nchen, \\
James-Franck-Stra{\ss}e, D-85747 Garching, Germany \\
Email: themmert@physik.tu-muenchen.de}

\maketitle

\abstracts{Recent developments 
in calculations of low energy nucleon properties utilizing effective
chiral field theories with explicit 
spin 3/2 matter fields are addressed.}
\section{Introduction}

The chiral symmetry of QCD is spontaneously broken at low energies, giving
rise to 3 [8] Goldstone Bosons for the case of 2 [3] light quark flavors.
In the following, we will concentrate on a world with only 
2 light flavors; all other
quark degrees of freedom are taken to be infinitely heavy. We identify the
3 Goldstone Boson degrees of freedom with the physical pions, which 
therefore owe their
small (but finite) mass to the additional explicit breaking of SU(2)$_L \times$
SU(2)$_R$ chiral symmetry due to the ``small'' masses of the up, down quarks.
Chiral Perturbation Theory is a successful effective field theory that
parameterizes the interactions among these Goldstone Bosons (in the presence of
external fields) in the most general form, based solely on the symmetries of
the underlying lagrangian of QCD. If this were all, this theory would obviously
be not very interesting for the audience of an NSTAR conference. However,
very general principles tell us also how these Goldstone bosons interact 
with ``matter''-fields, even in the presence of additional external
sources/fields. For an overview
on calculations involving Goldstone Bosons + matter fields  
I refer to the recent review of Ref.\cite{Ulf}.
Here, I will focus on ``matter'' that
consists of spin 1/2 fermions (Nucleons) and their spin 3/2 resonance partners
(Delta(1232)), constraining myself again to a world of 2 light flavors. In
particular, I will discuss the role of Delta resonances in microscopic
calculations of the anomalous magnetic moments of the Nucleon, the impact of
Deltas on the isovector Pauli form factor of the Nucleon and the problems
one faces if one wants to calculate the isovector Nucleon-Delta transition form
factors. Further topics of recent interest/activity, like the (reduced?)
screening of Delta(1232) generated paramagnetism in the {\em isoscalar}
Nucleon 
magnetic polarizability $\beta_M^{(s)}$ \cite{Harald} or the impact of 
Deltas on the momentum-dependence of the generalized spin-polarizabilities
of the Nucleon\cite{PRD}
cannot be covered here. 
\section{Chiral calculations and Power-counting}

For systematic calculations with chiral effective field theories one needs
a procedure to construct the most general effective chiral lagrangian that
contains all possible terms allowed by chiral symmetry, as well as (subsets of)
PCT constraints. In addition\footnote{Furthermore, one needs to specify, how
many light quark flavors one wants to consider, whether one utilizes
non-relativistic or relativistic chiral effective field theory and which
regularization procedure (consistent with the symmetries) one wants to 
employ.}, one needs to specify a power-counting scheme that
tells us, which ones of the plethora of possible diagrams and non-linear vertex
structures have to be taken into account if one performs a calculation up to 
a given order. In the following, I will show results obtained in 2 different 
chiral effective field theories: SU(2) HBChPT---which constitutes a 
non-relativistic
theory with only pion, Nucleon degrees of freedom and power-counting ${\cal
O}(p^n)$---and SU(2) SSE---which contains pion, Nucleon and Delta degrees of 
freedom and power-counting ${\cal O}(\epsilon^n)$. For details I have to refer
to the literature\cite{Ulf}.
 
Here, I only want to point out that in chiral effective field theories like
SSE\cite{SSE}, which contain the Nucleon-Delta mass splitting $\Delta_0$ as
an additional small parameter, there exists some degree of freedom as to how
one organizes the power-counting. The example I want to discuss here concerns
the leading $\gamma N\Delta$ vertex, which traditionally is written in the
(SU(6) quark-model inspired) form
\begin{eqnarray}
{\cal L}_{\gamma N\Delta}^{(2)}&=&\frac{b_1}{2 M_0}\bar{T}^\mu_j\;i\;
f_{+\mu\nu}^j\;S^\nu N_v +h.c.\;.
\end{eqnarray}
I do not want to discuss this operator structure in detail, the important point
to observe is that the dimension-less coupling $b_1$ is assumed to scale with a
large baryon mass scale $M_0$, boosting this structure to ${\cal
O}(\epsilon^2)$ (i.e. NLO) 
in the SSE lagrangian. Several reasons---for example the
resulting large value for the coupling $b_1\sim7.7$, the failure of SU(6)
symmetry considerations to predict the full strength of the $\gamma N\Delta$
transition, etc.---have compelled us to 
propose a slight modification for this operator:
\begin{eqnarray}
{\cal L}_{\gamma N\Delta}^{(1)}&=&\frac{c_V}{\Delta_0}\bar{T}^\mu_j\;i\;
f_{+\mu\nu}^j\;S^\nu N_v +h.c.\;.\label{vertex}
\end{eqnarray}
Due to the small mass scale $\Delta_0$ in the denominator, the {\em leading}
$\gamma N\Delta$-transition operator is now a part of the ${\cal
O}(\epsilon)$ (i.e. LO) SSE lagrangian, leading (in some cases)
to a substantial reordering of the chiral expansion. Interesting consequences
of this rescaled vertex of Eq.\ref{vertex} will be discussed in the following
sections. 
\section{Anomalous magnetic moments of the Nucleon}

\begin{figure}[t]
\centerline{\epsfig{file=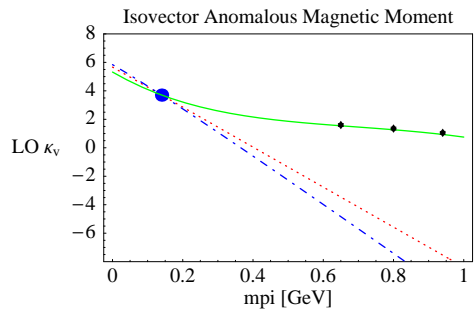,height=4cm}}
\caption[diag]{\label{kv} Isovector anomalous magnetic moment of the Nucleon
(in Nucleon magnetons) and its dependence on the effective mass of the pion.
The physical value at $m_\pi=140$ MeV is $\kappa_v=3.7$ [n.m]. Also shown are
3 (quenched) lattice QCD data points of Leinweber et al. \cite{Adelaide}; the
curves are explained in the text.}
\end{figure}

Figure \ref{kv} shows the leading one-loop order (LO) results obtained in 3 
different calculations utilizing chiral effective field theories for the 
anomalous isovector magnetic moment of the Nucleon $\kappa_v$ (in Nucleon
magnetons [n.m.]), defined via
\begin{eqnarray}
\kappa_v&=&\mu_{proton}-\mu_{neutron}-1\;[n.m.]
\end{eqnarray}
and plotted as a function of the pion mass $m_\pi$. 
The dotted curve shows the LO HBChPT
result of \cite{BKKM}
\begin{eqnarray}
\kappa_v^{(3)}&=&\kappa_v^{(0)}-\frac{g_A^2 M_N}{4\pi F_\pi^2}\;m_\pi
+{\cal O}(p^4)\; ,\label{linear}
\end{eqnarray}
with the free parameter $\kappa_v^0$ fixed in such a way that $\kappa_v^{(3)}$
reproduces the physical value of $\kappa_v^{(phys.)}=3.7$ [n.m.] at 
$m_\pi=140$ MeV. $g_A$ denotes the axial coupling constant of the Nucleon with
mass $M_N$, and $F_\pi$ is the pion-decay constant. All of these quantities are
taken at their physical values, as any implicit quark-mass dependence
constitutes an effect of higher order in the chiral expansion. Figure \ref{kv}
also shows that it is not possible to connect the LO HBChPT result with the
results of a quenched lattice QCD calculation by Leinweber et al. 
\cite{Adelaide}, employing effective pion masses of 600 MeV and higher.
A few years ago \cite{BFHM}, 
the LO influence of explicit Delta(1232) degrees of
freedom on the vector (and axial-vector) current of the Nucleon was analyzed
utilizing
the Small Scale Expansion of Ref.\cite{SSE}. As can be clearly seen by the 
dashed
curve in Figure \ref{kv}, there is only a small modification due to the
explicit spin 3/2 degrees of freedom for small pion mass. Still no 
satisfying connection to the lattice points can be obtained. 
\begin{figure}[t]
\centerline{\epsfig{file=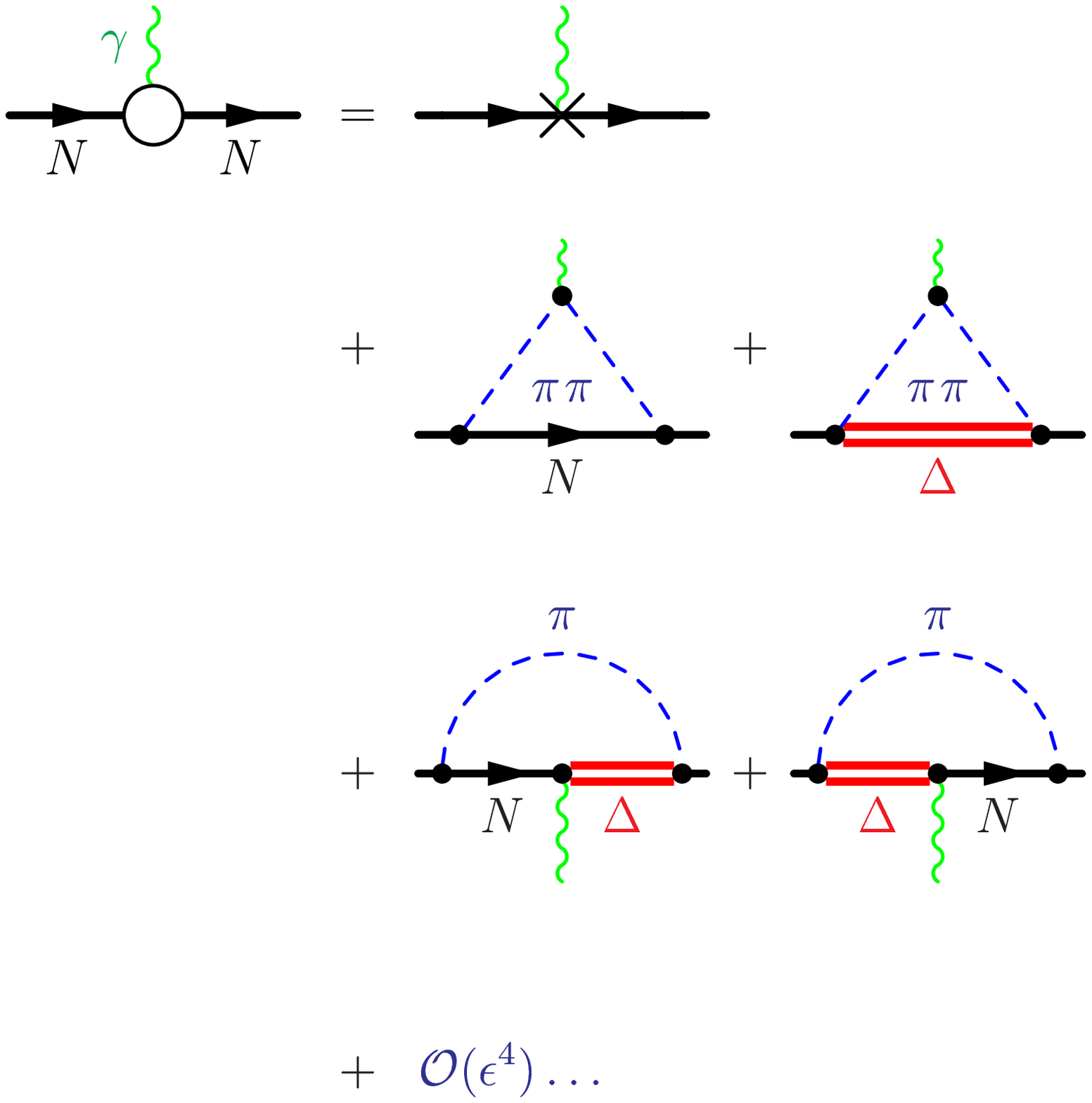,height=4cm}}
\caption[diag]{\label{diags} Diagrams contributing to the anomalous magnetic
moments of the Nucleon to leading one-loop order in the modified Small Scale
Expansion.}
\end{figure}

We now discuss the results of a new  calculation\cite{paper} of
the isovector anomalous magnetic moment which utilizes the rescaled leading 
$\gamma
N\Delta$ vertex of Eq.\ref{vertex} discussed in the previous section. The 
diagrams taken into account to leading one-loop order (i.e. 
${\cal O}(\epsilon^3)$) in the now modified Small Scale Expansion are shown in
Fig.\ref{diags}. The experts in the audience will notice that there are  2 
additional diagrams\footnote{We note that even the new set of one-loop
diagrams given in
Fig.\ref{diags} obtained to LO in the (modified) Small Scale Expansion contains
fewer diagrams than considered in the pioneering one-loop
calculation of the baryon
magnetic moments by Jenkins et al.\cite{Jenkins}. The 3 additional diagrams 
considered in Ref.\cite{Jenkins}
are part of the (estimated) 21 additional
one-loop diagrams coming in at NLO (i.e. ${\cal O}(\epsilon^4)$) 
in our approach, as
dictated by the SSE power-counting.} (in the last row) of Fig.\ref{diags} 
compared to the calculation of \cite{BFHM}. The result of this calculation
is shown by the full curve in Fig.\ref{kv}. It contains 3 free
parameters---$\kappa_v^0$, the isovector anomalous magnetic moment of the
Nucleon {\em in the chiral limit}, $c_V$, the new leading order $\gamma
N\Delta$ coupling constant introduced in Eq.\ref{vertex}, and one additional
(quark-mass dependent) higher order $\gamma NN$ coupling constant
$E_1(\lambda)$, which also serves as a counterterm to absorb new
divergences coming in due to the 2 extra diagrams. All other parameters can
be fixed from known low energy quantities. In Fig.\ref{kv}, we have fit
the 3 parameters
$\kappa_v^0,\,c_V,\,E_1(1\mbox{GeV})$ to the 3 (quenched) lattice QCD data
points of Leinweber et al.\cite{Adelaide}, but {\em not} to the $\kappa_v$
value at the physical pion mass. Surprisingly the fit---although
performed for pion masses $600 \ldots 950$ MeV !---suggests a quark-mass
dependence for the isovector anomalous magnetic moment of the Nucleon which
correctly
extrapolates down to $\kappa_v^{LO}=3.7$ at the physical pion mass of
$m_\pi=140$ MeV! Such a stable result obtained from only leading order input
could not be expected, especially if one compares it with the corresponding
HBChPT calculation (dotted curve). Details of the new calculation, 
including an error analysis,
will be available soon\cite{paper}. Here I, only want to point out that the 
analysis
suggests that the isovector anomalous magnetic moment has the chiral
limit value $\kappa_v^0\sim5.7$ [n.m.], which is more than 50\% larger than
the value for finite up, down quark masses. This dramatic reduction of the
anomalous magnetic moment is mainly due to 
pion-loop effects. Furthermore, one can
learn from Fig.\ref{kv} that the leading linear result of Eq.\ref{linear}
looses its range of validity already for values smaller than the physical pion
mass when curvature effects come in. This finding is consistent with a chiral
extrapolation of the same lattice QCD points analyzed with a Pade
approximation\cite{Tony}. Finally, I want to note\cite{paper} 
that the {\em isoscalar}
anomalous magnetic moment of the Nucleon shows quite a different/much simpler 
chiral behavior when calculated to the same order in SSE:
\begin{eqnarray}
\kappa_s^{LO}&=&\kappa_s^{(0)}-\frac{E_2 M_N}{4\pi^2 F_\pi^2\Delta}\;m_\pi^2
+{\cal O}(\epsilon^4)\; ,
\end{eqnarray} 
where $\kappa_s^0$ corresponds to the chiral limit value and $E_2$
denotes another unknown $\gamma NN$ coupling; the 2 parameters can
also be fitted to isoscalar lattice data, though with larger error 
bars due to the overall smallness of $\kappa_s$ \cite{paper}.
%
\section{Isovector Form Factors of the Nucleon}

\begin{figure}[t]
\centerline{\epsfig{file=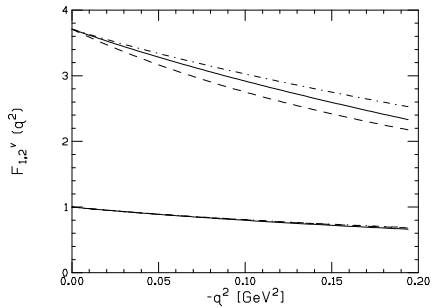,height=4cm}}
\caption[diag]{\label{f12v} Isovector Dirac and Pauli form factor of the
Nucleon, calculated to leading one-loop order with (full curve) and 
without (dot-dashed
curve) explicit spin 3/2 degrees of freedom. Also shown for comparison is a 
parameterization of the data\cite{Mergell} (dashed curve).}
\end{figure}

A few years ago, the role of explicit Delta(1232) degrees of freedom in 
calculations of the Nucleon form factors at low
four-momentum transfer  (i.e. $Q^2<0.2$ GeV$^2$)
was analyzed\cite{BFHM} within the Small Scale 
Expansion of Ref.\cite{SSE}. Considering the discussion in the previous 
section,
one can now ask the question how the results are changed (to leading one-loop
order) if one allows for the rescaling of the leading 
$\gamma N\Delta$-transition as defined in Eq.\ref{vertex}, which leads to the 
two additional (i.e. the last two) Feynman diagrams at  ${\cal O}(\epsilon^3)$
in Fig.\ref{diags} and to such a dramatic change in the LO chiral behavior
of the isovector anomalous magnetic moment as shown in Fig.\ref{kv}. However,
it turns out that to leading order the momentum-dependence
of the SSE curves of Ref.\cite{BFHM} is not modified by the additional 
diagrams for a physical pion mass $m_\pi=140$ MeV. 
The results for the isovector
Dirac and Pauli form factor of the Nucleon are shown in Fig.\ref{f12v}. 
The Dirac form factor turns out to be completely dominated by the radius in 
this
momentum-range. Both the HBChPT (${\cal O}(p^3)$, dot-dashed curve) and 
the (modified) SSE calculation 
(${\cal O}(\epsilon^3)$, full curve), as well as an empirical 
parameterization\cite{Mergell} of the data (dashed lines) cannot be 
distinguished. Matters
are different for the isovector Pauli form factor. Both the HBChPT and the
 SSE curves drop slower than suggested by the empirical
parameterization. However, the LO SSE calculation arising from the diagrams of
Fig.\ref{diags} provides an isovector
Pauli radius of 0.61 fm$^2$---arising solely from the pion-cloud around
a spin 1/2 or spin 3/2 intermediate baryon---which amounts to more than
75\% of the physical isovector Pauli radius and therefore leads to a
qualitatively better description of this form factor 
than provided by the LO HBChPT calculation of Ref.\cite{BKKM}
(cf. Fig.\ref{f12v}). It will be interesting to study the HBChPT-SSE 
comparison 
of the form factors also at NLO, to detemine how far in $q^2$ one
can trust/fine-tune chiral calculations at the one-loop level.  
\section{Isovector Nucleon-Delta Transition Form Factors}

\begin{figure}[t]
\centerline{\epsfig{file=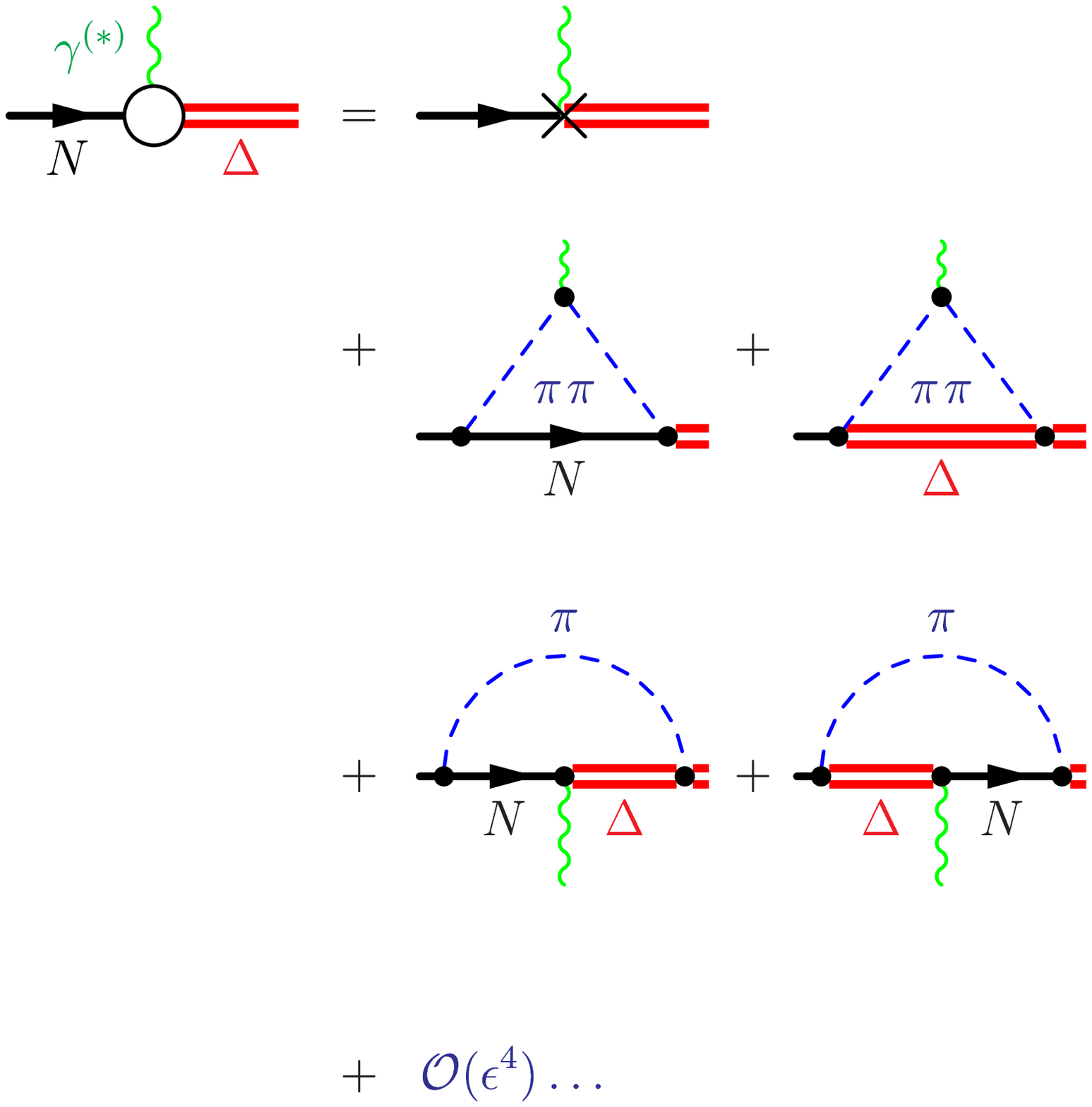,height=4cm}}
\caption[diag]{\label{gnd} Diagrams contributing to the isovector Nucleon-Delta
transition form factors to leading one-loop order in the modified Small Scale
Expansion.}
\end{figure}

\begin{figure}[t]
\centerline{\epsfig{file=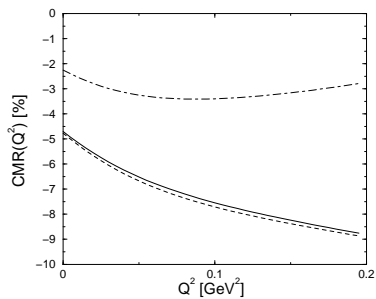,height=4cm}}
\caption[diag]{\label{cmr} Real part of the four-momentum dependence of 
the ratio of the C2/M1
$\gamma N\Delta$ transition moments, calculated to leading one-loop
order in the Small
Scale Expansion. The spread of the curves is discussed in the text.}
\end{figure}

Having discussed the impact of explicit Delta(1232) degrees of freedom in 
microscopic calculations of the (iso-) vector Nucleon current, one can now
address the analogous problems in the isovector Nucleon-Delta transition
current. This issue has already been analyzed within SSE in Ref.\cite{GHKP}.
Again, the rescaling of the leading order $\gamma N\Delta$ transition vertex
of Eq.\ref{vertex} leads to two additional diagrams (displayed in the last row
of Fig.\ref{gnd}), compared to the original calculation of Ref.\cite{GHKP}.
However, the additional diagrams only lead to a {\em 
modified chiral behavior} (i.e. a modified quark mass-dependence of
the $\gamma N\Delta$ transition moments\cite{paper}), whereas the 
{\em leading order}
four-momentum dependence for a
physical pion-mass of $m_\pi=140$ MeV
of the 3 isovector Nucleon-Delta transition form factors 
is unchanged. As discussed in Ref.\cite{GHKP}, microscopic calculations of
the $N\Delta$-transition 
using chiral effective field theories suffer (at present) from
the fact that there are several unknown counter-terms, which need to be fixed
via external input. As an example, in Fig.\ref{cmr}
we show the so called CMR, i.e. the ratio of
the C2 over the M1 Nucleon-Delta transition strength as a function of the
four-momentum transfer $q^2$. Depending on different ways of fixing the
counter-term input
as described in \cite{GHKP}, one obtains a rather broad band of possible
momentum-dependence for the CMR\footnote{For the status of experimental
information on the momentum dependence of CMR I refer to the talks by
R.W. Gothe, C.N Papanicolas, and H. Schmieden in these proceedings.}. 
The most promising way to improve upon these theoretical limitations
consists of a full calculation of the pion-electroproduction cross-section
{\em in the resonance region} within a chiral effective field theory like SSE.
So far, this has not been attempted\footnote{An exception is work of
Ref.\cite{Kao}, where the authors tried to utilize a low order SSE calculation
in the resonance region by averaging over the width of the resonance.} 
due to the high order required for an 
adequate treatment of the width of the Delta resonance. It is interesting to
note that the proposed rescaling
of the leading $\gamma N\Delta$ transition vertex Eq.\ref{vertex} 
also provides
new hope to get this longstanding problem in chiral effective field theory
calculations
finally done, as in the modified SSE the order required to include these width
effects is lowered; first exploratory studies are under way.

\section{Summary}

I have given a report regarding some new ideas/open problems 
in the field of chiral
effective field theories with explicit spin 3/2 resonance degrees of freedom.
I am convinced that the role and the treatment of baryon resonances continues 
to be an inspiring and challenging topic in this field. 

\begin{center}
{\bf Acknowledgments}
\end{center}

I want to thank the organizers of NSTAR01 for giving me the opportunity to
present these results to the community and I gratefully acknowledge their
financial support. I also would like to thank A. Thomas and W. Weise for 
helpful discussions. The research presented here was supported by BMBF. 


\begin{thebibliography}{99}
\bibitem{Ulf} see {\it e.g.} U.-G. Mei{\ss}ner, preprint no. 
{\tt [hep-ph/0007092]}. 
\bibitem{Harald} H.W. Grie{\ss}hammer and G. Rupak, preprint no. 
{\tt [nucl-th/0012096]}. 
\bibitem{PRD} see {\it e.g.} T.R. Hemmert, preprint no. 
{\tt [nucl-th/0101054]}.
\bibitem{SSE} T.R. Hemmert, B.R. Holstein and J. Kambor, J. Phys. {\bf G24}
(1998) 1831; Phys. Lett. {\bf B395} (1997) 89.
\bibitem{BKKM} V. Bernard et al., Nucl. Phys. {\bf B388}, 315 (1992).
\bibitem{BFHM} V. Bernard et al., Nucl. Phys. {\bf A635}, 121 (1998).
\bibitem{paper} T.R. Hemmert and W. Weise, forthcoming.
\bibitem{Jenkins} E. Jenkins et al., Phys. Lett. {\bf B302}, 482 (1993).
\bibitem{Adelaide} D.B. Leinweber, R.M. Woloshyn and T. Draper, Phys. Rev. 
{\bf D43}, 1659 (1991).
\bibitem{Tony} A.W. Thomas, these proceedings; see also D.B. Leinweber, D.H. Lu
and A.W. Thomas, Phys. Rev. {\bf D60}, 034014 (1999) and references therein.
\bibitem{Mergell} P. Mergell, U.-G. Mei{\ss}ner and D. Drechsel,
Nucl. Phys. {\bf A596}, 367 (1996).
\bibitem{GHKP} G.C. Gellas et al., Phys. Rev. {\bf D60}, 054022 (1999). 
\bibitem{Kao} C.W. Kao and T.D. Cohen, Phys. Rev. {\bf C60}, 064619 (1999). 
\end{thebibliography}
\end{document}